\documentstyle[twocolumn,pra,aps,psfig]{revtex}

\begin{document}
\title{Quantum noise in ideal operational amplifiers}
\author{Jean-Michel Courty, Francesca Grassia and Serge Reynaud
\thanks{courty, grassia or reynaud@spectro.jussieu.fr}}
\address{Laboratoire Kastler Brossel \thanks{
Laboratoire de l'Universit\'{e} Pierre et Marie Curie et de l'Ecole Normale
Sup\'{e}rieure associ\'{e} au Centre National de la Recherche Scientifique}, 
 case 74, \\ 4 place Jussieu, F-75252 Paris Cedex 05, France}
\date{November 1998}
\maketitle

\begin{abstract}
We consider a model of quantum measurement built on an ideal operational 
amplifier operating in the limit of infinite gain, infinite input impedance 
and null output impedance and with a feddback loop. 
We evaluate the intensity and voltage noises which have to be added 
to the classical amplification equations in order to fulfill the requirements 
of quantum mechanics. We give a description of this measurement device as a 
quantum network scattering quantum fluctuations from input to output ports.

\end{abstract}

Modern measurement techniques and especially ultrasensitive ones often
involve active systems. Such devices may be used either for a
preamplification purpose when a microscopic signal is amplified to an
observable macroscopic level, or for a stabilization purpose when a feedback
loop keeps the system in the vicinity of an optimal working point.
Amplifiers thus play a crucial role in ultrasensitive measurements and this
should be accounted for in theoretical analysis of ultimate sensitivities
attainable in quantum measurements.

Quantum noise associated with linear amplifiers has been the subject of
numerous works. In the line of thought initiated by early works on
fluctuation-dissipation relations \cite{Einstein05,Nyquist28} and continued
by a quantum analysis of linear response theory \cite{Callen51,Landau84},
active systems have been studied in the optical domain when maser and laser
amplifiers were developed \cite{Heffner62,Haus62,Gordon63}. General
thermodynamical constraints impose the existence of fluctuations for
amplification as well as for dissipation processes. At the limit of a null
temperature, these thermal fluctuations reduce to the quantum fluctuations
required by Heisenberg inequalities. This added noise determines the ultimate
performance of linear amplifiers \cite{Caves82,Loudon84} and plays a key
role in the question of optimal information transfer in optical
communication systems \cite{Gordon62,Takahasi65}. The theory of quantum
optical processes has led to the parallel development of a treatment of
quantum fluctuations which can be named as `quantum network theory' \cite
{Yurke84,Gardiner88}. It has been applied mainly to optical systems \cite
{Yamamoto90,Reynaud92} but it has also been viewed as a generalized quantum
extension of the linear response theory which is of interest for electrical
systems as well \cite{Courty92}.

Most practical applications of amplifiers in measurements involve ideal
operational amplifiers operating near the limits of infinite gain, infinite
input impedance and null output impedance. It could appear difficult to deal
with these limits without pathologies in the treatment of fluctuations.
In the present letter we show how this difficulty may be circumvented.

To this aim, we study a model of quantum measurement performed with an 
ideal operational amplifier. The model, sketched on Figure 1, consists
in an ideal operational amplifier operating with a feedback loop and 
connecting two coaxial lines denoted $l$ and $r$ for left and right. 
These two lines will be associated respectively with signal and readout.
The presence of a feedback loop fixes the effective gain and effective 
impedances of the device. It entails that a third line $f$ has to be
introduced to account for the fluctuations associated with the
dissipative part of the feedback impedance. 
We show in the letter that the ultimate performances of this measurement 
device may be characterized in a precise manner. 
Fluctuations added by the amplifier are described in 
terms of noise spectra of equivalent voltage and current generators. 
Since these fluctuations account for quantum as well as thermodynamical 
constraints, they are described by non commuting operators and obey Heisenberg
inequalities.

\begin{figure}[tbp]
\centerline{\psfig{figure=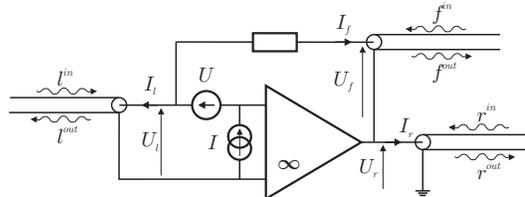,width=7cm}}
\caption{Sketch of the quantum measurement apparatus built on an ideal
operational amplifier.}
\end{figure}

Free fields are propagating in the inward and outward directions in each line
coupled to the amplifier. The left line comes from a monitored electrical system
so that the inward field $l^{\rm in}$ plays the role of the signal to be
measured. Meanwhile, the right line goes to an electrical meter so
that the outward field $r^{\rm out}$ is the meter readout. In connection
with the discussions of Quantum Non Demolition measurements in quantum
optical systems \cite{Grangier92,Braginsky92}, $l^{\rm out}$ appears as
the back-action field sent back to the monitored system and $r^{\rm in}$
represents the fluctuations coming from the readout line. The feedback loop
is partly dissipative and then contains a Nyquist noise source. This Nyquist
noise is an extra inward field $f^{\rm in}$ coming through a coaxial line $%
f$ representing the feedback resistance. The outward field $f^{\rm out}$
represents the way the dissipated energy leaves the system. In principle,
information on the measurement can be extracted through this output channel
but it is usually lost and the line $f$ is only a noise source in this case.

We now present the electrical equations associated with the measurement
device of Fig.1. We first write the characteristic relations between the
various voltages and currents 
\begin{eqnarray}
U &=&U_{l}=U_{r}+U_{f}+\left( Z_{f}-R_{f}\right) I_{f}  \nonumber \\
I &=&I_{l}+I_{f}  \label{amplifier}
\end{eqnarray}
Here, $U_{p}$ and $I_{p}$ are the voltage and current at the port $p$, {\it i.e.}
at the end of the line $p=l$, $r$ or $f$ while $U$ and $I$ are the voltage
and current noise generators associated with the operational amplifier
itself. $Z_{f}$ is the impedance of the feedback loop and $R_{f}$
its dissipative part, {\it i.e.} the characteristic impedance of the line $l$
so that $\left( Z_{f}-R_{f}\right) $ is the reactive impedance of the
feedback loop. All equations are implicitly written in the frequency
representation and the impedances are functions of frequency. Equations (\ref
{amplifier}) take a simple form because of the limits of infinite gain,
infinite input impedance and null output impedance assumed for the
operational amplifier. They correspond to the limit of a more general
treatment which has been given elsewhere with a finite gain as well as
finite input and output impedances \cite{Grassia98}.

As a second step, we rewrite the voltage and current $U_{p}$ and $I_{p}$ at
the port $p$ in terms of the inward and outward fields $p^{\rm in}$ and $%
p^{\rm out}$ counterpropagating in the line $p$ which has a characteristic
impedance $R_{p}$ 
\begin{eqnarray}
I_{p} &=&\sqrt{\frac{\hbar \left| \omega \right| }{2R_{p}}}\left( p^{\rm out}
-p^{\rm in}\right)  \nonumber \\
U_{p} &=&\sqrt{\frac{\hbar \left| \omega \right| R_{p}}{2}}\left( p^{\rm out}
+p^{\rm in}\right)  \label{lines}
\end{eqnarray}
The coaxial line being equivalent to a one-dimensional space, the input
fields $p^{\rm in}$ may be described by the quantum theory of free fields 
in a two-dimensional space-time. In particular,
they obey the standard commutation relations of such a theory 
\begin{equation}
\left[ p^{\rm in}\left[ \omega \right] ,p^{\rm in}\left[ \omega ^{\prime
}\right] \right] =2\pi \ \varepsilon \left( \omega \right) \ \delta \left(
\omega +\omega ^{\prime }\right)  \label{freecommut}
\end{equation}
where $\varepsilon \left( \omega \right) $ denotes the sign function. This
relation just means that the positive and negative frequency components
correspond respectively to the annihilation and creation operators of
quantum field theory. Fields corresponding to different lines commute with
each other.

The fluctuations of these fields will then be characterized by a noise
spectrum $\sigma _{pp}^{\rm in}$ with its well-known expression for a
thermal equilibrium at a temperature $T_{p}$ 
\begin{eqnarray}
\left\langle p^{\rm in}\left[ \omega \right] \cdot p^{\rm in}\left[
\omega ^{\prime }\right] \right\rangle &=&2\pi \ \sigma _{pp}^{\rm in}
\left[ \omega \right] \ \delta \left( \omega +\omega ^{\prime }\right) 
\nonumber \\
\sigma _{pp}^{\rm in}\left[ \omega \right] &=&\frac{1}{2}\coth \frac{\hbar
\left| \omega \right| }{2k_{B}T_{p}}  \label{thermal}
\end{eqnarray}
The dot symbol denotes a symmetrized product and $k_{B}$ is the
Boltzmann constant. The quantity $\hbar \left| \omega \right| \sigma _{pp}^
{\rm in}$ is the energy per mode. It reduces to the zero point energy $%
\frac{\hbar \left| \omega \right| }{2}$ at the limit of zero temperature and
to the classical result $k_{B}T_{p}$ at the high temperature limit. We also
assume that the fields incoming through the various ports are uncorrelated
with each other as well as with amplifier noises.

Now the output fields $p^{\rm out}$ have their fluctuations determined by
the transformation of the input fields through their interaction with the
measurement device. This is the common idea of all input-output descriptions
of quantum networks 
\cite{Yurke84,Gardiner88,Yamamoto90,Reynaud92,Courty92,Grangier92}.
After this transformation, the field fluctuations are no longer 
described by the thermal correlation functions given in (\ref{thermal}). 
Moreover, fluctuations are no longer independent in different ports.
However the output fields $p^{\rm out}$ still obey the
commutation relations (\ref{freecommut}) of free fields \cite{Courty92}. In
the present letter we use this property to characterize the quantum
fluctuations of the voltage and current sources associated with the
operational amplifier.

To this aim, we use the characteristic equations (\ref{amplifier},\ref{lines}%
) associated with the amplifier and the lines to rewrite the output fields $%
l^{\rm out}$, $r^{\rm out}$ and $f^{\rm out}$ in terms of input fields 
$l^{\rm in}$, $r^{\rm in}$, $f^{\rm in}$ and of amplifier noise
sources $U$ and $I$ 
\begin{eqnarray}
l^{\rm out} &=&-l^{\rm in}+\sqrt{\frac{2}{\hbar \left| \omega \right|
R_{l}}}U  \nonumber \\
r^{\rm out} &=&-r^{\rm in}-2\frac{Z_{f}}{\sqrt{R_{r}R_{l}}}l^{\rm in}-2%
\sqrt{\frac{R_{f}}{R_{r}}}f^{\rm in}  \nonumber \\
&+&\sqrt{\frac{2}{\hbar \left| \omega \right| R_{r}}}\left( \frac{R_{l}+Z_{f}%
}{R_{l}}U-Z_{f}I\right)   \nonumber \\
f^{\rm out} &=&f^{\rm in}+2\sqrt{\frac{R_{f}}{R_{l}}}l^{\rm in}+\sqrt{%
\frac{2R_{f}}{\hbar \left| \omega \right| }}\left( I-\frac{U}{R_{l}}\right) 
\label{inout}
\end{eqnarray}
Knowing that the output fields obey the same commutation relations (\ref
{freecommut}) as the input fields and that they commute with voltage and
current fluctuations $U$ and $I$, we deduce from (\ref{inout}) that the
latter obey the commutation relations of conjugate observables 
\begin{eqnarray}
\left[ U\left[ \omega \right] ,U\left[ \omega ^{\prime }\right] \right] 
&=&\left[ I\left[ \omega \right] ,I\left[ \omega ^{\prime }\right] \right] =0
\nonumber \\
\left[ U\left[ \omega \right] ,I\left[ \omega ^{\prime }\right] \right] 
&=&2\pi \ \hbar \omega \ \delta \left( \omega +\omega ^{\prime }\right) 
\label{UU}
\end{eqnarray}
In fact the two first relations in (\ref{inout}) are sufficient to
demonstrate (\ref{UU}) and the consistency with the third one may then be
verified. Commutators (\ref{UU}) entail that voltage and current
fluctuations verify Heisenberg inequalities which determine the ultimate
performance of the ideal operational amplifier used as a measurement device.

To push this analysis further it is worth introducing new quantities 
$a^{\rm in}$ and $c^{\rm in}$ as linear combinations of the noises $U$ and 
$I$ depending on a factor $R$ having the dimension of an impedance 
\begin{eqnarray}
\sqrt{2\hbar \left| \omega \right| }a^{\rm in} &=&\sqrt{R}I+\frac{U}{\sqrt{R}}
\nonumber \\
\sqrt{2\hbar \left| \omega \right| }c^{\rm in} &=&\sqrt{R}I-\frac{U}{\sqrt{R}}
\label{defac}
\end{eqnarray}
For an arbitrary value of $R$, the quantities $a^{\rm in}$ and $c^{\rm in}$ 
obey the following commutation relations 
\begin{eqnarray}
\left[ a^{\rm in}\left[ \omega \right] ,a^{\rm in}\left[ \omega ^{\prime
}\right] \right] &=&-\left[ c^{\rm in}\left[ \omega \right] ,c^{\rm in}
\left[ \omega ^{\prime }\right] \right] 
\nonumber \\
&=&2\pi \ \varepsilon \left( \omega
\right) \ \delta \left( \omega +\omega ^{\prime }\right) 
\nonumber \\
\left[ a^{\rm in}\left[ \omega \right] ,c^{\rm in}\left[ \omega ^{\prime
}\right] \right] &=& 0 
\end{eqnarray}
This means that $a^{\rm in}$ can be interpreted as the input field
in a new line $a$ and $c^{\rm in}$ as the field conjugated to the input 
field $b^{\rm in}$ in another new line $b$ 
\begin{eqnarray}
c^{\rm in}\left[ \omega \right]  &=&b^{\rm in}\left[ -\omega \right]  
\nonumber \\
\sigma _{cc}^{\rm in}\left[ \omega \right]  &=&\sigma _{bb}^{\rm in}
\left[ \omega \right]   \nonumber \\
\left[ b^{\rm in}\left[ \omega \right] ,b^{\rm in}\left[ \omega ^{\prime
}\right] \right]  &=&2\pi \ \varepsilon \left( \omega \right) \ \delta
\left( \omega +\omega ^{\prime }\right)   \label{defbc}
\end{eqnarray}
In other words, the voltage and current noises associated with the amplifier
may be replaced by the coupling to $2$ further lines $a$ and $b$ and the
presence of amplification requires a conjugation of fluctuations coming in
one of these two lines. Conjugation means here a change of sign for frequencies 
or, equivalently, an exchange of annihilation and creation operators. 
This latter feature was already known for linear amplifiers 
\cite{Caves82,Loudon84}.

We may then fix the parameter $R$ to a value $R_{0}$ chosen so that the
fields $a^{\rm in}$ and $b^{\rm in}$ are uncorrelated fluctuations. It
follows from (\ref{defac}) that this specific value is determined by the
ratio between voltage and current noise spectra 
\begin{equation}
R_{0}=\sqrt{\frac{\sigma _{UU}}{\sigma _{II}}}  \label{R0}
\end{equation}
The $2$ noise spectra $\sigma _{UU}$ and $\sigma _{II}$ are defined as
symmetric correlation functions as in (\ref{thermal}). The fields $a^{\rm in}$ 
and $b^{\rm in}$ are thus described by temperatures $T_{a}$ and $T_{b}$
as in (\ref{thermal}). We have implicitly assumed that these fluctuations
are the same for all field quadratures, i.e. that the amplifier noises are
phase-insensitive. We may also consider for simplicity that the 
value $R_{0}$ is constant over the spectral domain of interest
although this assumption is not mandatory for the forthcoming analysis.
It is worth emphasizing that the fluctuations of $U$ and $I$ deduced
from (\ref{defac}) are generally correlated. The only case where
they are uncorrelated corresponds to noise temperatures equal in the two 
lines $a$ and $b$.

\begin{figure}[tbp]
\centerline{\psfig{figure=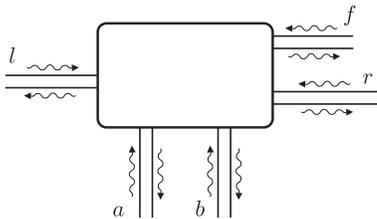,width=5cm}}
\caption{Representation of the measurement apparatus of Fig.1
as a $5$-port quantum network.}
\end{figure}

The amplifier of Figure 1 is now depicted as the $5$-port quantum network of
Figure 2 which couples the $5$ dissipation lines $l$, $r$, $f$, $a$, $b$. The
advantage of this new picture is that all noise sources are now attributed 
to free fields coming through the dissipation lines and are thus 
separated from the purely reactive elements gathered in the network. 
As a consequence, the transformation of fields by the reactive
network is described by a unitary $5 \times 5$ scattering matrix. We give
below the resulting expressions for the output fields $l^{\rm out}$ and $%
r^{\rm out}$ in the signal and readout lines 
\begin{eqnarray}
l^{\rm out} &=&-l^{\rm in}+\sqrt{\frac{R_{0}}{R_{l}}}\left( a^{\rm in}
-c^{\rm in}\right)   \nonumber \\
r^{\rm out} &=&-r^{\rm in}-2\frac{Z_{f}}{\sqrt{R_{r}R_{l}}}l^{\rm in}-2%
\sqrt{\frac{R_{f}}{R_{r}}}f^{\rm in}  \nonumber \\
&&+\left( 1+\frac{Z_{f}}{R_{l}}-\frac{Z_{f}}{R_{0}}\right) \sqrt{\frac{R_{0}%
}{R_{r}}}a^{\rm in}  \nonumber \\
&&-\left( 1+\frac{Z_{f}}{R_{l}}+\frac{Z_{f}}{R_{0}}\right) \sqrt{\frac{R_{0}%
}{R_{r}}}c^{\rm in}  \label{InOut}
\end{eqnarray}
The first relation describes the back-action noise induced by the
measurement device on the signal line. Here, this noise is just the voltage 
noise source associated with the amplifier. We recall
that $c^{\rm in}$ is in fact the conjugate of an input field $b^{\rm in}$
defined exactly as the other ones (see (\ref{defbc})). We do not discuss 
the back-action noise in more detail and concentrate the forthcoming
discussion on the second relation which describes the added noise.
Notice that equations (\ref{InOut}) also allow to study the correlations between 
the output fields $l^{\rm out}$ and $r^{\rm out}$ or between these fields 
and the input ones. The terms complementing the unitarity scattering
matrix may be found in \cite{Grassia98}.

In order to characterize the performance of the measurement device in terms
of added noise, we introduce an estimator $\widehat{l}^{\rm in}$ of the
signal $l^{\rm in}$ as it may be deduced from the knowledge of the meter
readout $r^{\rm out}$ 
\begin{eqnarray}
\widehat{l}^{\rm in} &=&-\frac{\sqrt{R_{r}R_{l}}}{2Z_{f}}r^{\rm out} 
\nonumber \\
&=&l^{\rm in}+\frac{\sqrt{R_{l}}}{Z_{f}}\left( \frac{\sqrt{R_{r}}}{2}
r^{\rm in}+\sqrt{R_{f}}f^{\rm in}\right)   \nonumber \\
&&-\frac{1}{2}\left( \frac{1}{Z_{f}}+\frac{1}{R_{l}}-\frac{1}{R_{0}}\right) 
\sqrt{R_{l}R_{0}}a^{\rm in}  \nonumber \\
&&+\frac{1}{2}\left( \frac{1}{Z_{f}}+\frac{1}{R_{l}}+\frac{1}{R_{0}}\right) 
\sqrt{R_{l}R_{0}}c^{\rm in}  \label{estim}
\end{eqnarray}
The estimator $\widehat{l}^{\rm in}$ would be identical to the measured
signal $l^{\rm in}$ in the absence of added fluctuations. Hence the noise
added by the measurement device is described by the supplementary terms
assigned respectively to Nyquist noises $r^{\rm in}$ and $f^{\rm in}$ in
the readout and feedback lines as well as Nyquist noises $a^{\rm in}$ and 
$c^{\rm in}$ in the two lines representing amplification noises. Notice that
proper fluctuations of $l^{\rm in}$ are included in the signal and not in the
added noise. 

The whole added noise is characterized by a spectrum obtained as a
sum of the uncorrelated noise spectra associated with these Nyquist noises 
\begin{eqnarray}
\Sigma  &=&\frac{R_{l}R_{r}}{4\left| Z_{f}\right| ^{2}}\sigma _{rr}^{\rm in}
+\frac{R_{l}R_{f}}{\left| Z_{f}\right| ^{2}}\sigma _{ff}^{\rm in} 
\nonumber \\
&&+\frac{R_{l}R_{0}}{4}\left| \frac{1}{Z_{f}}+\frac{1}{R_{l}}-\frac{1}{R_{0}}%
\right| ^{2}\sigma _{aa}^{\rm in}  \nonumber \\
&&+\frac{R_{l}R_{0}}{4}\left| \frac{1}{Z_{f}}+\frac{1}{R_{l}}+\frac{1}{R_{0}}%
\right| ^{2}\sigma _{bb}^{\rm in}  \label{sigma}
\end{eqnarray}
The Nyquist spectra are given by thermal equilibrium relations (\ref{thermal}%
) with temperatures $T_{r}$, $T_{f}$, $T_{a}$ and $T_{b}$. The expression (%
\ref{sigma}) allows us to evaluate the added noise for arbitrary values
of the impedance parameters and of these noise temperatures. In particular
the amplification noises are characterized by the three parameters $R_0$,
$T_{a}$ and $T_{b}$. 

When concentrating the discussion on ultimate sensitivity of the
measurement apparatus, we see from (\ref{sigma}) that it is wise to
have the feedback impedance large enough so that only the effects of
amplifier noises persist. Introducing the parameter 
\begin{equation}
e^{\xi }=\sqrt{\frac{R_l}{R_0}} \label{defxi}
\end{equation}
we rewrite (\ref{estim},\ref{sigma}) under the simple forms
\begin{eqnarray}
\widehat{l}^{\rm in} &=&l^{\rm in}+\sinh \xi \,a^{\rm in}+\cosh \xi
\,c^{\rm in}  \nonumber \\
\Sigma  &=&\sinh ^{2}\xi \, \sigma _{aa}^{\rm in}+ \cosh ^{2}\xi 
\, \sigma_{bb}^{\rm in}  
\label{Sigma}
\end{eqnarray}
If the two temperatures $T_{a}$ and $T_{b}$ are fixed, the added noise is
made minimal by choosing the parameter $\xi$ equal to zero, that is also 
by matching the values of the two impedance parameters $R_{0}$ and $R_{l}$.
Notice that this is not a good solution for decreasing the back-action
noise (see (\ref{InOut})). But this is the optimum as far as the criterium
of minimizing added noise is privileged. Then, relations (\ref{Sigma}) are 
read as 
\begin{eqnarray}
\widehat{l}^{\rm in} &=&l^{\rm in} + c^{\rm in}  
\nonumber \\
\Sigma  &=&\sigma_{bb}^{\rm in} 
\label{SigmaMin}
\end{eqnarray}
Finally this added noise is still decreased by going to a temperature
$T_{b}$ as low as possible. At the limit of a null temperature, we
recover the optimum of $3\,{\rm dB}$ added noise which is 
the same as for phase-insensitive linear amplifiers \cite{Takahasi65}.

In this letter, we have studied the quantum noise associated with an ideal
operational amplifier used as a phase-insensitive measurement apparatus. 
We have given a description of this apparatus as a quantum network coupling
$5$ lines associated with the signal and readout lines, with the feedback
resistance and with the two lines representing voltage and current noises
added by the amplifier. We have obtained an expression of the noise
added by the measurement which depends on the various impedance parameters
and on the corresponding noise temperatures. We have shown that 
the ultimate performance of the device is reached when the following 
conditions are met: large feedback impedance $Z_f$, impedance $R_l$ of the 
signal line matched to the parameter $R_0$ characterizing the amplification
noises, null temperature for these amplification noises. 

As argued in the Introduction, amplifiers play an important 
role in most real-life high-sensitivity measurements. 
Amplification with an infinite gain may be considered as the archetypal 
description of the transition from a microscopic quantum signal to
a macroscopic classical readout \cite{Osawa97}.
Amplifiers are also involved in active stabilization techniques.
The representation of the ideal operational amplifier as a quantum network
helps to treat it as an element used in more sophisticated systems. 
Hence, the results presented in the present letter should open the way to 
a renewed analysis of quantum measurements with active devices. 

\noindent
{\bf Acknowledgements}  We thank Vincent Josselin, Fulvio Ricci, 
Pierre Touboul and Eric Willemenot for fruitful discussions.

\end{document}